\documentclass[sigconf]{cidr-2027}

\makeatletter
\def\subsubsection{\@startsection{subsubsection}{3}%
 \z@{.5\linespacing\@plus.7\linespacing}{.1\linespacing}%
 {\it\sf\bf}}
\makeatother

\renewcommand\footnotetextcopyrightpermission[1]{}

\usepackage{multirow}
\usepackage{tcolorbox}
\usepackage{balance} 
\usepackage{booktabs} 
\usepackage{amsmath}
\usepackage{graphicx}
\usepackage{tikz}
\usetikzlibrary{calc,arrows.meta,shapes.geometric,fit}
\usepackage{csquotes}
\usepackage{array}
\usepackage{cuted}
\usepackage[linesnumbered,ruled,vlined]{algorithm2e}
\SetCommentSty{\color{blue!70}}
\usepackage{epstopdf}

\usepackage{afterpage}
\usepackage{makecell}
\usepackage{listings}
\usepackage[normalem]{ulem}
\definecolor{codegreen}{rgb}{0,0.6,0}
\definecolor{codegray}{rgb}{0.5,0.5,0.5}
\definecolor{codepurple}{rgb}{0.58,0,0.82}
\definecolor{backcolour}{rgb}{0.95,0.95,0.92}
\newcommand{\myline}[1]{{\medskip\noindent\textbf{#1.}}}

\usepackage{url}
\usepackage{enumerate}
\usepackage[labelfont=bf]{caption}
\usepackage{xcolor}
\usepackage{color, colortbl}

\usepackage[utf8]{inputenc}
\usepackage{amsthm}
\usepackage[linesnumbered,ruled,vlined]{algorithm2e}

\SetKw{ParallelForEach}{parallel \ForEach}

\usepackage{listings}
\usepackage{pifont}

\lstdefinestyle{mystyle}{
  backgroundcolor=\color{backcolour},  
  commentstyle=\color{codegreen},
  keywordstyle=\color{magenta},
  numberstyle=\tiny\color{codegray},
  stringstyle=\color{codepurple},
  basicstyle=\ttfamily\small,
  breakatwhitespace=false,     0.00510417
  breaklines=true,         
  captionpos=b,          
  keepspaces=true,         
  numbersep=5pt,         
  showspaces=false,        
  showstringspaces=false,
  showtabs=false,
  otherkeywords = {<<, >>, WITH}, 
  tabsize=2
}

\makeatletter
\def\@ACM@checkaffil{}
\makeatother

\begin{document}
\pagestyle{plain} 


\title{Git4Data: Database-Native Version Control for AI Agents}

\author{Hongshen Gou \;\; Zuyu Zhang\;\; Yuze Sun \;\; Peng Xu \;\; Feng Tian \;\; Long Wang \;\; Jianguo Wang$^\S$}
\affiliation{
\vspace{0.1cm}
 \institution{\textsf{MatrixOrigin} \;\; \textsf{Purdue University}$^\S$}
 \vspace{0.1cm}
  \city{\textit{\{gouhongsheng; zuyuzhang; sunyuze; xupeng\}@matrixorigin.cn} \;\; \{tianfeng; wanglong\}@matrixorigin.io \;\; \textit{csjgwang@purdue.edu}$^\S$}
}







\begin{abstract}
Large Language Model (LLM) agents increasingly explore many candidate states of relational data in parallel, each of which should remain isolated, reproducible, and auditable, preferably through the same SQL interface used for ordinary data work. Existing tools support this requirement only partially: source-code version control does not scale to large datasets, whereas relational databases manage large data efficiently but rarely expose native branching, comparison, and merging. We present Git4Data, a database-native version-control layer for agentic workflows. Git4Data treats a database as a repository and a table as a versioned object, exposing Git-style operations (\texttt{snapshot}/\texttt{tag}, \texttt{branch}, \texttt{diff}, and \texttt{merge} with explicit conflict-resolution policies) through SQL extensions. Implemented in MatrixOne, a cloud-native relational database, Git4Data leverages immutable object storage and MVCC to make the cost of these operations proportional to the size of the change rather than the size of the data. On the BranchBench agentic branching workloads, Git4Data outperforms DoltDB by up to an order of magnitude. Overall, we believe this work sheds light on how relational databases can better support AI agents through efficient versioning.


\end{abstract}


\maketitle

\section{Introduction}\label{sec:intro}


Large Language Model (LLM) agents are beginning to act as data
engineers~\cite{LLMDisrupt}: they read relational data, propose
transformations~\cite{FMWrangle}, evaluate SQL~\cite{NL2SQLDawn}, and iterate. Unlike a human engineer working on one
snapshot at a time, a fleet of agents explores many candidate states in
parallel~\cite{AgentFirstDataSystems,BranchBench}, and each agent must be
isolated, reproducible, and auditable. An agent must fork data before
speculative updates, inspect row-level diffs, merge only validated changes,
and roll back failed paths without copying the base version. In other words, agentic workflows
require data-version primitives as first-class, database-native
operations rather than as external tools.

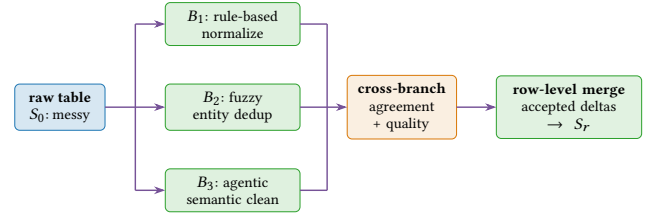
\begin{figure}[t]
  \centering
  \definecolor{MainDataColor}{HTML}{1F77B4}
  \definecolor{BranchColor}{HTML}{2CA02C}
  \definecolor{EvalColor}{HTML}{9467BD}
  \definecolor{RejectColor}{HTML}{D62728}
  \definecolor{DiffColor}{HTML}{D97706}
  \resizebox{0.98\linewidth}{!}{%
  \begin{tikzpicture}[
    font=\scriptsize,
    box/.style={
      draw,
      rounded corners=2pt,
      align=center,
      inner sep=3pt,
      outer sep=0pt,
      minimum height=0.52cm,
    },
    main/.style={box, draw=MainDataColor, fill=MainDataColor!18, text=black},
    branch/.style={
      box,
      draw=BranchColor,
      fill=BranchColor!15,
      text=black,
      text width=1.7cm,
      minimum width=1.9cm,
    },
    diff/.style={box, draw=DiffColor, fill=DiffColor!12, text=black},
    merge/.style={
      box,
      draw=BranchColor,
      fill=BranchColor!14,
      text=black,
      text width=1.85cm,
      minimum width=1.85cm,
    },
    prearrow/.style={
      -{Stealth[length=1.6mm,width=1.1mm]},
      line width=0.6pt,
      draw=EvalColor!80!black,
    },
    wsarrow/.style={
      -{Stealth[length=1.6mm,width=1.1mm]},
      line width=0.58pt,
      draw=EvalColor!80!black,
    },
    wsbus/.style={
      line width=0.58pt,
      draw=EvalColor!80!black,
    },
    wsline/.style={
      line width=0.58pt,
      draw=EvalColor!80!black,
    },
  ]
    \node[main, minimum width=1.28cm] (raw) at (0,0)
      {\textbf{raw table}\\$S_0$: messy};

    \node[branch] (b1) at (2.45,1.18) {$B_1$: rule-based\\normalize};
    \node[branch] (b2) at (2.45,0) {$B_2$: fuzzy\\entity dedup};
    \node[branch] (b3) at (2.45,-1.18) {$B_3$: agentic\\semantic clean};

    \node[diff, minimum width=1.34cm, text width=1.34cm] (score) at (4.85,0)
      {\textbf{cross-branch}\\agreement\\+ quality};
    \node[merge] (merge) at (7.22,0)
      {\textbf{row-level merge}\\accepted deltas\\$\rightarrow S_r$};

    \coordinate (split) at (1.07,0);
    \coordinate (join) at (3.79,0);
    \coordinate (splitTop) at (1.07,1.18);
    \coordinate (splitBot) at (1.07,-1.18);
    \coordinate (joinTop) at (3.79,1.18);
    \coordinate (joinBot) at (3.79,-1.18);
    \draw[wsline] (raw.east) -- (split);
    \draw[wsbus] (splitTop) -- (splitBot);
    \draw[wsarrow] (splitTop) -- (b1.west);
    \draw[wsarrow] (split) -- (b2.west);
    \draw[wsarrow] (splitBot) -- (b3.west);
    \draw[wsbus] (joinTop) -- (joinBot);
    \draw[wsline] (b1.east) -- (joinTop);
    \draw[wsline] (b2.east) -- (join);
    \draw[wsline] (b3.east) -- (joinBot);
    \draw[wsarrow] (join) -- (score.west);
    \draw[wsarrow] (score.east) -- (merge.west);
  \end{tikzpicture}%
  }
  \caption{Agentic data repair workflow with Git4Data}
  \label{fig:agentic_workflow}
\end{figure}



Consider agentic data repair workflow in Figure~\ref{fig:agentic_workflow}. A raw
snapshot \(S_0\) contains duplicate entities, malformed fields, and
inconsistent records. Rather than committing to one cleaning strategy up
front, an agent branches \(S_0\) and applies a different SQL repair to each
candidate: categorical normalization, fuzzy entity deduplication, and
context-dependent semantic fixes. Each evaluates the candidates with SQL queries
and proxy quality signals, prunes poor branches, and compares promising ones
through diffs. Because each strategy repairs a different subset of the
records, no branch wins outright, and the goal is to selectively merge accepted
row-level deltas into a repaired state \(S_r\) while preserving an audit trail
of what was tried, accepted, and rejected.

Software engineering already has mature support for these primitives, with
Git~\cite{Git} as the de facto standard. However, data engineers do not have a comparable foundation. Applying a source-code version control system (VCS) to large
datasets is prohibitively slow, because its diff and merge tools load and
compare the entire dataset in memory, and it does not scale to billions of records stored in cloud object stores.


Relational databases mutate data efficiently via transactions, but do not provide explicit \emph{version control}.
Multi-Version Concurrency Control (MVCC) retains row versions in a linear history, but does not allow tracing back at a specific version,
let alone doing so at agent swarm scale. Database snapshots and Point-In-Time Recovery (PITR) capture named past versions, yet support only read and
restore at a single timeline, but cannot hold two writable lines of work. Writable clones of tables or databases in Snowflake~\cite{Snowflake} and Supabase~\cite{Supabase} allow development and testing in parallel, and Neon~\cite{Neon}
forks an entire Postgres instance at any point in its history through
storage-level copy-on-write. But none of these systems can report the row-level differences between two branches, whether both branches modify the same rows, or merge validated changes into a new version. In other words, what is missing in databases is not the ability to capture or fork data changes, but the SQL-level operations to manipulate the resulting versions: comparison and conflict-aware reintegration.

We propose \emph{Git4Data}, a new abstraction that supplies exactly this missing layer. Git4Data treats a database as a repository and a table as a versioned
object. By extending snapshot/tag, branch, row-level diff, and merge as SQL statements (Section~\ref{sec:vcop}), an engineer or an LLM agent could version data through the interface already used for data work, and changes spanning multiple tables publish in one transaction.
Merge is three-way and conflict-aware rather than last-writer-win. 
Conflict resolution is currently at the row granularity, and
richer semantic resolution is one of the research problems this abstraction opens
(Section~\ref{sec:lessons}).

We then implement an initial version of Git4Data in MatrixOne~\cite{MatrixOne},
a cloud-native relational database. Our key observation is that a modern OLTP
database already provides the mechanisms for the concept, where data resides in immutable, append-only objects governed by MVCC. A table version is just a lightweight metadata,
and two versions differ only in the objects written since they diverged,
letting diff and merge read only those \emph{deltas}. The \textbf{initial results are
encouraging}: cloning a 100~GB table takes 0.2~s and a few hundred kilobytes of
metadata; diff and merge outperform their SQL equivalents by orders of
magnitude. On BranchBench~\cite{BranchBench} agentic branching workloads Git4Data runs up to an order of
magnitude faster than DoltDB~\cite{Dolt} while sustaining 1{,}000 concurrently branching agents.


This paper makes the following contributions: 
\begin{itemize}
\item We propose \emph{Git4Data}, a database-native version-control
abstraction that exposes branch, diff, and merge as SQL extensions, applicable to any OLTP database (Section~\ref{sec:vcop}).
\item We implement a Git4Data prototype in MatrixOne\footnote{Git4Data is open source as part
of MatrixOne at \url{https://github.com/matrixorigin/matrixone}.}, where metadata-only cloning and delta-based diff and merge
keep the cost proportional to the changes 
(Section~\ref{sec:vcopimpl}).
\item We evaluate the prototype on
BranchBench, observing up to an order of magnitude
speedup over DoltDB at 1{,}000 concurrent agents
(Section~\ref{sec:eval}).
\item We distill the lessons of building Git4Data and identify the open
problems that
we believe start a new line of research on in-database version
control (Section~\ref{sec:lessons}).
\end{itemize}



\section{Git4Data}\label{sec:vcop}
This section presents Git4Data, which derives much of its utility from a small, composable vocabulary in Git: 
record a version, branch from it, compare versions, and integrate selected changes.
Git4Data brings this vocabulary to relational data by mapping each Git concept
onto a database construct. 
The database serves as the data repository where each table is a versioned object:
\begin{itemize}
\item a \emph{snapshot} names an immutable table state, the analogue of a commit or tag;
\item a \emph{branch} is a new table cloned from a snapshot, after which the two evolve independently;
\item \emph{diff} reports the rows on which two versions disagree;
\item \emph{merge} folds accepted changes from one branch into another under an explicit conflict policy.
\end{itemize}

Two decisions shape how the vocabulary transfers. First, each operation is a
SQL statement executed inside the database, so versioning inherits the
transactions, authentication, and access control that data engineers already
rely on, and both engineers and agents version data through the same
interface they use for ordinary data work. Second, versions are compared as
relational content rather than as byte streams, where Git4Data treats a table version as an unordered multiset of rows, with the
primary key, when present, supplying a stable row identity across versions, so
the semantics are independent of physical layout and row order.

Nothing in this vocabulary is tied to one engine: the operations are defined at
the SQL level, and any OLTP database could achieve them with the same semantics.
Their cost, however, depends on the storage design. If the storage engine keeps table
data immutable, a snapshot is merely metadata: branching does not occur data copies, and two versions differ only in what were written after they diverged,
so each operation costs in proportion to the change rather than to the whole table size.

\subsection{Operations}
We present the operations in their typical workflow order, as a single
table \texttt{T} takes a snapshot or a branch, compares, and reconciles.
Listing~\ref{lst:wf} depicts the running workflow used throughout the
paper, where \texttt{T} and its clone \texttt{TClone} diverge from a common base
snapshot \texttt{sn1}, and eventually merge into \texttt{sn4}.

\begin{lstlisting}[label=lst:wf,caption=Running branch-and-merge workflow,basicstyle=\small\ttfamily]
  T:       --> sn1 --> sn2 ------>  sn4 -->
                \             /
  TClone:        \---> sn3 --/

         ------------- now  ---------> time
\end{lstlisting}

\myline{Snapshot}
Every version-control workflow requires the ability to name a past state. In
Git4Data, a snapshot freezes a table at an instant, and plays the role of a Git
commit. The most lightweight form is implicit: a multi-version storage engine already
retains point-in-time history for a recent window (i.e., 24~hours), so a recent state can be queried directly by timestamp,

\begin{verbatim}
  SELECT * FROM T{timestamp='2026-08-01 12:34:56'};
\end{verbatim}
without requiring the user to declare the state in advance. When a state should
be retained explicitly, the user promotes it to a \emph{named} snapshot, the
analogue of a Git tag, with 
\begin{verbatim}
  CREATE SNAPSHOT sn1 FOR TABLE T;
\end{verbatim}
We write $T_{sn1}$ for the resulting snapshot of \texttt{T}. Git4Data also
supports database-level snapshots; for clarity we develop the operations at the
granularity of a single table.

\myline{Branch}
A snapshot becomes a starting point for new work the moment it is cloned into a
fresh table:
\begin{verbatim}
  DATA BRANCH CREATE TABLE TClone FROM T{snapshot='sn1'};
\end{verbatim}
The clone \texttt{TClone} inherits the schema and data of $T_{sn1}$, but from that
point the two tables evolve independently: inserts, deletes, and updates on
\texttt{T} and \texttt{TClone} no longer affect one another. This is precisely the
isolation an agent needs to explore a speculative change without endangering
production state. In the running workflow, \texttt{T} then advances to snapshot
\texttt{sn2} while \texttt{TClone} advances to \texttt{sn3}.

\myline{Diff}
Once two lines of work diverge, the natural question is how they differ.
\texttt{DATA BRANCH DIFF} compares between two snapshots:
\begin{verbatim}
  DATA BRANCH DIFF T{snapshot='sn2'} AGAINST TClone{snapshot='sn3'};
\end{verbatim}

Conceptually, the operation treats each snapshot as an unordered multiset of
records and reports the rows where the two multisets disagree.
The semantics are captured precisely by the query in
Listing~\ref{lst:diff}: each row contributes a signed count, and the rows whose
counts do not cancel are exactly those that differ.


\begin{lstlisting}[label=lst:diff,caption=Query equivalent to DATA BRANCH DIFF,basicstyle=\small\ttfamily]
  WITH UnionT AS (
    SELECT -1 AS cnt, a, b, c FROM T{snapshot='sn2'}
    UNION ALL
    SELECT 1 AS cnt, a, b, c FROM TClone{snapshot='sn3'}
  )
  SELECT SUM(cnt) AS diffCnt, a, b, c FROM UnionT
  GROUP BY a, b, c HAVING SUM(cnt) <> 0;
\end{lstlisting}


\myline{Merge}
This operation folds accepted changes back into the target table. Branches from a fleet of agents
typically succeed on different subsets of the data, so no single branch can
simply be promoted to replace the original. Meanwhile the live table keeps
receiving writes, as shown in Listing~\ref{lst:wf}, \texttt{T} advances to \texttt{sn2}
while \texttt{TClone} is being explored, so swapping \texttt{T} for a branch
would silently discard that concurrent progress, and re-applying the branch's
changes through hand-written SQL reintroduces a slow, error-prone path. A merge reconciles two histories instead:
\begin{verbatim}
  DATA BRANCH MERGE TClone{snapshot='sn3'} INTO T
    [WHEN CONFLICT FAIL|SKIP|ACCEPT];
\end{verbatim}
The source may be any snapshot, whereas the target must be the live table
version. Rather than overwriting one side with the other, Git4Data infers the
common base revision $T_{sn1}$ and performs a three-way merge, so that
non-overlapping changes from both branches survive. What happens when the branches
do overlap is governed by the \texttt{WHEN CONFLICT} clause, which offers three
policies: \texttt{FAIL} aborts the merge, \texttt{SKIP} keeps the target's version
of a conflicting row, and \texttt{ACCEPT} keeps the source's.

\myline{Conflict Resolution}
Whether two changes actually collide depends on how rows are identified, and here
the presence of a primary key is decisive. When the table declares one, Git4Data
compares the corresponding row across the base, target, and source snapshots and
flags a genuine conflict only when both branches independently modify the same
key, including the case in which two branches insert the same new key. If only one
branch touched the row, or if both applied identical changes, the outcome is
unambiguous and Git4Data resolves it automatically.

Without a primary key, no stable identity ties a row in one branch to a row in
another, so Git4Data falls back to multiset reasoning. Inserted rows are grouped
by their full values, while deleted rows are tracked through the storage engine's physical row
identifiers. A conflict is suspected only when the
same changed row or row value appears in both branch deltas and cannot be
canceled as an identical change; changes confined to one side are applied
automatically, and anything that remains is resolved under \texttt{SKIP},
\texttt{ACCEPT}, or \texttt{FAIL}.

\section{Design and Implementation}\label{sec:vcopimpl}
In this section, we present the design and implementation of Git4Data. 
We first identify the
capabilities that an OLTP database must provide to support it efficiently,
then describe MatrixOne, a cloud-native database,
and finally explain how each operation is implemented.

\subsection{Key Requirements}\label{sec:requirements}
Git4Data operations rely on three requirements, which we state independently of any particular storage engine.

\myline{Append-only data}
The storage engine never modifies data in place: inserts and updates append new,
immutable data units.
A snapshot only needs to record the table version to which the units belong,
branching copies the metadata, and the difference between two versions is
confined to the units appended after they diverged.

\myline{Deletion marks}
Deleted tuples must be recorded explicitly, rather than applied in place.
Deletion marks allow diff to report deleted rows without scanning the full
table, while merge can distinguish a row deleted on one branch from a row
deleted on both.

\myline{Multi-Version Concurrency Control}
Version-control operations must execute as transactions, so that a
merge publishes its accepted changes atomically and concurrent branches never
observe a partial state. MVCC additionally
gives every committed state a well-defined point-in-time identity, 
the implicit timestamp of snapshot name.

Many modern OLTP databases built on log-structured, multi-version storage engine
satisfy all three requirements. We implement Git4Data in
MatrixOne~\cite{MatrixOne}, one such system described below.

\subsection{MatrixOne}\label{sec:matrixone}

\begin{figure}[t!]
  \centering
   \resizebox{0.8\linewidth}{!}{%
  \begin{tikzpicture}[
    font=\small,
    cn/.style={draw, semithick, ellipse, minimum width=1.2cm,
      minimum height=0.78cm, inner sep=1pt},
    tn/.style={draw, semithick, ellipse, minimum width=1.7cm,
      minimum height=1.05cm},
    log/.style={draw, semithick, cylinder, shape border rotate=90,
      aspect=0.35, minimum width=1.5cm, minimum height=0.8cm,
      inner sep=1.5pt, font=\scriptsize},
    store/.style={draw, semithick, minimum height=0.75cm},
    arr/.style={-{Stealth[length=1.8mm,width=1.4mm]}, semithick},
    biarr/.style={{Stealth[length=1.8mm,width=1.4mm]}-%
      {Stealth[length=1.8mm,width=1.4mm]}, semithick},
  ]
    \node[cn] (cn0) at (0.5, 0) {ComputeNode0};
    \node[cn] (cn1) at (3.5, 0) {ComputeNode1};
    \node[cn] (cn2) at (6.5, 0) {ComputeNode2};
    \draw[semithick] (-1.2,-1.0) -- (8.25,-1.0);
    \foreach \i in {0,1,2} {
      \draw[arr] (cn\i.south |- {(0,-1.0)}) -- (cn\i.south);
    }
    \node[tn] (tn) at (5.5,-2.45) {TrxNode};
    \draw[biarr] (tn.north) -- (tn.north |- {(0,-1.0)});
    \node[log] (ls2) at (1.85,-2.15) {LogService2};
    \node[log] (ls1) at (0.75,-3.15) {LogService1};
    \node[log] (ls3) at (2.95,-3.15) {LogService3};
    \node[draw, semithick, inner sep=7pt, fit=(ls1)(ls2)(ls3)] (raft) {};
    \node[anchor=north west, inner sep=2.5pt] at (raft.north west) {Raft};
    \draw[arr] (tn.west) -- (raft.east |- tn.west);
    \node[store, minimum width=9.7cm] (os) at (3.55,-4.35)
      {Cloud Object Storage};
    \draw[arr] (tn.south) -- (tn.south |- os.north);
    \coordinate (busR) at (7.5,-1.0);
    \draw[biarr] (busR) -- (busR |- os.north);
  \end{tikzpicture}%
  }
  \caption{MatrixOne architecture}
  \Description{MatrixOne Database Architecture}
  \label{fig:m1a}
\end{figure}
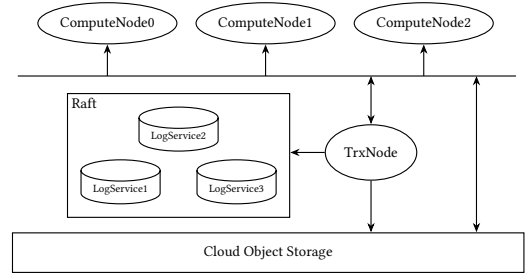

MatrixOne is a cloud-native HTAP database organized around three principal node types (Figure~\ref{fig:m1a}).
LogService nodes form a Raft~\cite{Raft} group and store the Write-Ahead Log
(WAL). A TransactionNode (TN) determines transaction commits, serializes committed
logs, and streams WAL records to subscribed ComputeNodes (CNs). CNs execute
SQL queries and scale out independently.

MatrixOne stores table data in cloud object storage. Objects are
immutable, hold column-store row groups~\cite{ColumnStore}, and form an LSM
tree~\cite{LSMTree} ordered by the primary key, or by clustering keys together
with a uniquifier. Deletes are represented by tombstone objects that record the
key and physical row id of the deleted rows. Metadata consists of a
directory of data objects and tombstone objects.

MatrixOne employs MVCC. Transactions
execute on CNs and maintain private workspaces. Large transaction workspaces
are written directly to cloud object storage, and the commit record sent to TN carries
the corresponding object metadata. Small writes may instead accumulate in a
TN-managed in-memory object; each row carries a transaction timestamp, and the
object remains append-only until flushed to cloud. Committed WAL
records are streamed to subscribed CNs to keep table state
and local object caches fresh.

A table snapshot is the metadata directory for all objects that belong to the
table at a given version, including both flushed in-memory objects and
remote objects. Reading a timestamp snapshot traverses this directory and
applies MVCC timestamp filters. Creating a named snapshot first flushes in-memory
objects and then records the metadata directory under the snapshot name.

As in other log-structured storage systems~\cite{LSMTree,LSFS}, MatrixOne relies
on background compaction and garbage collection. Garbage collection is
snapshot-aware: objects referenced by named snapshots are retained, which allows
Git4Data to maintain branches and tags without copying table data.

\subsection{Implementation}
We now describe how Git4Data implements clone, diff, and merge over
MatrixOne snapshots. Cloning a table from a snapshot is pure metadata: MatrixOne
copies the directory structure of the snapshot's object metadata into the new
table.

Diff and merge operate on object deltas. Consider tables \texttt{T} and
\texttt{TClone} in Listing~\ref{lst:wf}: after \texttt{TClone} is created, both
may be modified independently, advancing to distinct snapshots \texttt{sn2} and
\texttt{sn3}. We write $\Delta_{sn2}$ for the objects in $T_{sn2}$ but not in the
common base revision $T_{sn1}$, and $\Delta_{sn3}$ analogously for $TClone_{sn3}$.
A delta contains the objects added by data modifications, with deletions
represented as tombstone records. It may also contain objects retained only in
$T_{sn1}$ as a result of compaction or garbage collection.

\subsection{Diff}\label{sec:diff}
To compute \texttt{DATA BRANCH DIFF} between $T_{sn2}$ and $TClone_{sn3}$,
MatrixOne reads only $\Delta_{sn2}$ and $\Delta_{sn3}$. With a primary key, all
operations on one key within a delta collapse into a single logical operation: a
delete, an insert, or an update represented as a delete followed by an insert.
Deletions are always applied to a row in $T_{sn1}$. This scan mirrors an ordinary
LSM-tree scan with tombstones, except that it emits deletions rather than masking deleted rows,
assigning a minus sign to each deletion and a plus sign to each insert. It differs
from the SQL diff of Listing~\ref{lst:diff} in two ways: its signs express
$T_{sn2}$ versus $T_{sn1}$ (not versus $TClone_{sn3}$), and deleted rows initially
emit only tombstones whose non-key columns are null. MatrixOne joins with
$T_{sn1}$ to recover the original values only when they are required.

MatrixOne then aggregates the two deltas: changes are identical, and cancel, when
they delete the same row of $T_{sn1}$ or insert rows equal in all columns. The
final output flips the sign of rows from $\Delta_{sn2}$ and, for each remaining
tombstone, joins with $T_{sn1}$ to recover its non-key columns. Without a primary
key, the same procedure applies, but rows are matched by full value and by
physical rowid: insertions with identical values and deletions with the same rowid
cancel, and deleted rows are recovered by rowid lookup.

\subsection{Merge}\label{sec:merge}
A three-way merge from \texttt{TClone} into \texttt{T} runs the same scan and diff
aggregation over $\Delta_{sn2}$ and $\Delta_{sn3}$ (Section~\ref{sec:diff}); the
plus and minus signs indicate whether a row was inserted into a snapshot or
deleted from the base. With a primary key, a conflict is genuine when the key
appears in both deltas and spurious when it appears on only one side; a row
relocated by compaction (same values, new position) is recognized as a
non-conflict, so a storage reorganization never masks a valid update from the
other branch. This is the only case requiring a full read of the deleted base row,
and such lookups are rare. Without a primary key, rows are matched as a multiset of
full values: identical full-row values from the two deltas cancel one occurrence
at a time, preserving duplicate multiplicity, so equal rows on both sides are
cancellations rather than conflicts, consistent with the row-multiset semantics of
Section~\ref{sec:vcop}.

Users need not name the common base revision. MatrixOne tracks snapshot and clone
lineage and usually infers it, implementing a two-way merge as a three-way merge
with an implicit base. When the base cannot be determined or no longer exists
(for example, the original table and all of its snapshots were deleted), the merge
uses an empty base; even then, two clones of a common ancestor share many objects,
so the diff aggregation still beats the SQL query of Listing~\ref{lst:diff} by
skipping the shared objects.

Snapshots also protect history: MatrixOne never compacts or garbage-collects
objects referenced by a named snapshot. Compaction or GC scheduled between
\texttt{sn1} and a later snapshot rewrites valid rows into new objects, which can
move rows without changing their values; as above, the diff aggregation treats
such relocated rows as unchanged and avoids false conflicts. Because users
typically branch from well-organized snapshots, compaction within the common base
revision is rare.


\section{Experimental Evaluation}\label{sec:eval}
This section evaluates Git4Data on a microbenchmark
and compares against DoltDB~\cite{Dolt} on
BranchBench~\cite{BranchBench}. Our experiments were conducted on a bare-metal server running CentOS,
equipped with an Intel Xeon Silver CPU (2.4~GHz, 64 cores), 256~GB of main
memory, and local SSDs, with MatrixOne deployed on Kubernetes.

\subsection{Version-Control Operations}\label{sec:eval_ops}
We first microbenchmark the individual version-control operations on both
primary-key (PK) and no-primary-key (NoPK) settings. We compare Git4Data
against equivalent hand-written SQL in MatrixOne using the TPC-H~\cite{TPCH}
\texttt{lineitem} table at Scale Factor~100.

\myline{Clone}
We compare the metadata-only clone against materializing a full copy of the
\texttt{lineitem} table:
\begin{verbatim}
  INSERT INTO T SELECT * FROM lineitem;
\end{verbatim}
Table~\ref{tab:clone} reports the cost of clone versus insert. Cloning from a
snapshot copies only metadata, whereas the \texttt{INSERT} writes an entirely new table, incurring 34~GB of additional storage.

\begin{table}[tb]
  \centering
  \caption{Git4Data Clone vs.\ Insert on 100~GB \texttt{lineitem} table.}
  \label{tab:clone}
  \begin{tabular}{lcc}
    \toprule
    \textbf{Operation} & \textbf{Time (s)} & \textbf{Space} \\
    \midrule
    Clone, PK  & 0.20 & 314 KB \\
    Clone, NoPK & 0.17 & 294 KB \\
    Insert, PK  & 114.6 & 34 GB \\
    Insert, NoPK & 119.3 & 34 GB \\
    \bottomrule
  \end{tabular}
\end{table}

\myline{Diff and merge}
We apply four update sets on the \texttt{lineitem} clones:
C1, C2, C3, and C4 update 1{,}000, 10{,}000, 100{,}000, and 1{,}000{,}000 random rows, respectively, and then differ against the original table and
merge the changes back using \texttt{ACCEPT}. The SQL diff shows in
Listing~\ref{lst:diff}, while the SQL merge materializes that diff, deletes the
rows with negative counts, and inserts the rows with positive counts.

\begin{table}[tb]
  \centering
  \caption{Single-branch diff, Git4Data vs.\ SQL.}
  \label{tab:expdiff}
  \begin{tabular}{lcccc}
    \toprule
    \textbf{Operation} & \textbf{C1} & \textbf{C2} & \textbf{C3} & \textbf{C4} \\
    \midrule
    Git4Data Diff, PK  & 0.19 & 0.38 & 1.73 & 3.27 \\
    Git4Data Diff, NoPK  & 0.85 & 22.50 & 9.04 & 60.19 \\
    SQL Diff, PK  & 316.16 & 418.19 & 428.78 & 431.50 \\
    SQL Diff, NoPK  & 378.19 & 396.61 & 394.13 & 371.74 \\
    \bottomrule
  \end{tabular}
\end{table}

The built-in diff is substantially faster than the SQL diff
(Table~\ref{tab:expdiff}), because it scans only the changed objects rather than
the entire tables. The advantage is larger with primary keys, since row ids
let MatrixOne collapse multiple operations on the same key before aggregation.
In the no-primary-key case, deleted rows may require additional tuple lookups, so
the benefit is smaller and varies with the different number of lookups.

\begin{table}[tb]
  \centering
  \caption{Single-branch merge, Git4Data vs.\ SQL.}
  \label{tab:expmerge}
  \begin{tabular}{lcccc}
    \toprule
    \textbf{Operation} & \textbf{C1} & \textbf{C2} & \textbf{C3} & \textbf{C4} \\
    \midrule
    Git4Data Merge, PK  & 0.35 & 0.97 & 7.95 & 16.13 \\
    Git4Data Merge, NoPK  & 0.88 & 22.70 & 12.09 & 68.75 \\
    SQL Merge, PK  & 321.52 & 412.89 & 442.68 & 471.16 \\
    SQL Merge, NoPK  & 393.95 & 405.35 & 401.59 & 403.18 \\
    \bottomrule
  \end{tabular}
\end{table}

The built-in merge is likewise substantially faster than the SQL alternative
(Table~\ref{tab:expmerge}). Primary keys provide stable row identity, allowing
the scan and diff aggregation to group multiple operations on the same key into a
single logical operation.

The same advantage holds in collaborative settings. For four engineers who fork
\texttt{lineitem} and merge mostly non-overlapping updates, with two branches
conflicting on a 10\% PK overlap resolved by \texttt{ACCEPT}, the built-in diff
and merge remain orders of magnitude faster than the SQL counterparts, even for
one million updates.

\subsection{BranchBench}\label{sec:eval_agentic}
We now turn to end-to-end agentic workloads,
studying how Git4Data behaves as agent count and data size grow.

BranchBench~\cite{BranchBench} defines branch lifecycle, branch-local SQL,
cross-branch comparison, and pruning workloads for agentic database branching. We
run four end-to-end workflows, software development, failure reproduction, data
cleaning, and Monte Carlo tree search (MCTS), at scale factor~100
(approximately 47 million rows). Each is driven by five concurrent agents over 20
steps, with each agent forking the database, issuing branch-local SQL, and merging
or discarding its branch. We compare Git4Data against
DoltDB~\cite{Dolt}, a MySQL-compatible database with built-in Git-style branching,
reporting end-to-end wall-clock time for a cold run and the average over warm runs
in Table~\ref{tab:branchbench}. A fifth workflow, \texttt{simulation}, stresses
branch concurrency with 1{,}000 agents and is examined separately below.


\begin{table}[tb]
  \centering
  \caption{BranchBench at Scale Factor~100 runtime (s)}
  \label{tab:branchbench}
  \setlength{\tabcolsep}{4pt}
  \begin{tabular}{lccccc}
    \toprule
    & \multicolumn{2}{c}{\textbf{Git4Data}} & \multicolumn{2}{c}{\textbf{DoltDB}} & \\
    \cmidrule(lr){2-3}\cmidrule(lr){4-5}
    \textbf{Workflow} & \textbf{Cold} & \textbf{Warm} & \textbf{Cold} & \textbf{Warm} & \textbf{Speedup} \\
    \midrule
    \texttt{software\_dev}  & 138.9 & 122.1 & 1938.8 & 1925.6 & 15.8$\times$ \\
    \texttt{failure\_repro} & 207.6 & 198.9 & 1545.5 & 1677.3 & 8.4$\times$ \\
    \texttt{data\_cleaning} & 62.6  & 58.6  & 1075.2 & 1084.2 & 18.5$\times$ \\
    \texttt{mcts}           & 36.0  & 39.8  & 410.4  & 410.2  & 10.3$\times$ \\
    \bottomrule
  \end{tabular}
\end{table}

Across all four workflows, Git4Data is up to $18.5\times$ faster than DoltDB,
which materializes and compares table contents, so each branch operation scales with table size.
Git4Data's snapshot-and-delta operations scale with the size of the change,
and
exhibit low runtime variance under 3.5~s. 

\myline{Scaling number of agents}
The \texttt{simulation} workflow stresses branch concurrency directly, launching
1{,}000 concurrent agents, each of which forks the database and performs
a branch-local step. At scale factor 100, Git4Data completes in 400~s, whileas DoltDB fails to finish within two hours.
This confirms that metadata-level branching keeps per-fork cost negligible
even at high agent counts. The remaining cost is dominated not by branching itself
but the shared compute and I/O consumed when thousands of branch-local
executions run concurrently, which we identify as the primary bottleneck as the
number of agents grows.


\begin{table}[tb]
  \centering
  \caption{Git4Data BranchBench scaling experiments.}
  \label{tab:scaling}
  \setlength{\tabcolsep}{6pt}
  \begin{tabular}{lccc}
    \toprule
    \textbf{Workflow} & \textbf{SF100} & \textbf{SF1,000} & \textbf{Factor} \\
    \midrule
    \texttt{software\_dev}  & 127.5 & 366.3  & 2.9$\times$ \\
    \texttt{data\_cleaning} & 99.3  & 322.5  & 3.2$\times$ \\
    \texttt{failure\_repro} & 199.0 & 2685.7 & 13.5$\times$ \\
    \bottomrule
  \end{tabular}
\end{table}

\myline{Scaling data size}
To see how Git4Data performs as data grow, we rerun the BranchBench
workflows at scale factor~1{,}000 (Table~\ref{tab:scaling}). The branch-local workflows
scale sublinearly: \texttt{software\_dev} and \texttt{data\_cleaning} slow by at
most $3.2\times$, and \texttt{mcts} stays essentially flat (under 40~s warm), even
though the data grows by $10\times$. Their cost is set by the bounded per-step
deltas each agent writes, not by the size of the underlying table. The only
exception is \texttt{failure\_repro}, which grows about $13\times$: its repairs
scan and rewrite the entire table, so its cost tracks data size rather than change
size. The 1{,}000-agent \texttt{simulation}
finishes in 600~s at SF1{,}000, comparable to 400~s at SF100, which confirms that per-fork cost stays metadata-bound as data grows.
Overall, Git4Data's delta-oriented design preserves its efficiency at an
order-of-magnitude larger scale, except where a workflow inherently touches the
whole table.

\section{Related Work}\label{sec:related}
This section reviews prior work in five dimensions: version control, storage
snapshots, data-lake versioning, database cloning and branching, and agentic data workloads.


Version control systems trace back to early software-engineering
practice~\cite{SCCS}. Distributed systems such as Git~\cite{Git}
subsequently became the standard for source-code collaboration. Support for
large data is typically retrofitted as an extension: Git LFS~\cite{GitLFS} provides 
large-file storage and retrieval, and DVC~\cite{DVC} versions datasets
and models through Git-tracked pointer files, but neither offers record-level
diff or conflict resolution.


Closer to Git4Data in spirit are data-lake versioning systems. Apache
Iceberg~\cite{Iceberg} maintains per-table snapshot lineage with branch and tag
references, Nessie~\cite{Nessie} versions an entire Iceberg catalog with
Git-like branches and merges, and lakeFS~\cite{LakeFS} provides zero-copy
branches, commits, and three-way merge over an object-store namespace. These
systems share Git4Data's metadata-only branching philosophy: a lakeFS commit,
for example, is an immutable manifest whose merge copies unchanged ranges
wholesale. They version file manifests, but the unit of diff identity is an
object or a table rather than a row: two
branches that update disjoint rows of the same table still collide, and a
conflict is resolved by keeping one side's file. Git4Data applies the same
snapshot-and-delta discipline inside the OLTP database, where primary keys give
rows a unique identity and merge reconciles at record granularity.

Many database systems support snapshots, restore, and Point-In-Time Recovery
(PITR), but creating a \emph{writable} branch exposes the limits of engines
that couple compute with in-place-update storage. PostgreSQL~18 can
delegate the database copy to file-system cloning (reflink)~\cite{PGCreateDB}, but requires that no
other active session access the source database for the duration of the copy,
making it effectively an offline operation. Disaggregated, versioned storage removes the copy
altogether: Snowflake~\cite{Snowflake} and Supabase~\cite{Supabase} offer
zero-copy clones for development and testing, and Neon~\cite{Neon} branches
online, at no cost to the parent, by recording a fork point in its LSN-indexed
page history. Neon's branches, however, are coarse-grained: a single table cannot
branch alone. Moreover, none of these systems can compare two branches at
the row level or merge one into another, so divergence is one-way.
DoltDB~\cite{Dolt} brings Git-style branch, diff, and merge operations to a
MySQL-compatible SQL database. Git4Data likewise exposes branch-oriented data
workflows, but realizes them as database-native snapshot operations at table
granularity in MatrixOne, with an emphasis on efficient diff and merge over
large table forks.

Agentic workloads provide a further motivation for branchable data management, as they
frequently require parallel exploration of a hypothetical data space. Recent work
on agent-first data systems characterizes this pattern as agentic speculation~\cite{AgentFirstDataSystems}: agents may fork a database state, run speculative updates, and roll back
branches. Each branch furnishes an isolated state where an agent can apply speculative mutations, assess their effects, and then
discard, compare, or merge the resulting changes. BranchBench~\cite{BranchBench}
is a recent benchmark for agentic database branching that models branch lifecycle
operations, branch-local SQL, cross-branch comparison, and pruning as first-class
workload dimensions (Section~\ref{sec:eval}).

\section{Lessons and Future Work}\label{sec:lessons}

We share three lessons learned from building Git4Data that we believe extend beyond MatrixOne, and finally highlight important open problems.


\myline{Data version control is a storage property}
Our implementation shows that data version control can be built on top of an
existing transactional storage engine. An engine that already provides the
three capabilities of Section~\ref{sec:requirements}, append-only data,
deletion marks, and transactions, contains the necessary mechanisms: Git4Data
is a thin interpretation layer over them and required no modification to the
storage layer itself. We hope this work opens a line of research on data
version management inside databases.

\myline{Relational semantics simplify data versioning}
While Git matches ordered lines of text heuristically, a relational engine obtains cleaner semantics for free.
Data reconciliation based on primary key index outperforms the traditional value-based matching by more than an order of magnitude. But diff and merge assume compatible schemas, so schema evolution on long-lived branches
remains open. 

\myline{New challenges surface in scaling the agent swarm}
Once forking is a metadata operation and merges commit atomically, branch creation no longer constrains agent concurrency, and the dominant cost becomes the shared computation and I/O of concurrently executing branch-local workloads, shifting the open problem from storage efficiency to resource governance.

\myline{Future directions}
The bottleneck shift defines the first direction: \emph{resource governance for
agent fleets}. Thousands of speculative branches compete for shared compute and
I/O, most will be discarded, yet the engine treats them as equal tenants;
scheduling, admission control, and per-branch quotas that reflect an agent's
progress are open problems. Second, \emph{richer merge semantics}. Resolution
is currently row-level, so two branches editing different columns of the same
row still conflict; cell-level resolution is the natural next step, but the
harder problem is semantic, since row-disjoint changes can jointly violate
constraints that neither branch violates alone, and conflict policies beyond
\texttt{SKIP} and \texttt{ACCEPT}, including an agent that reviews \texttt{DATA
BRANCH DIFF} output and acts as the merge driver, are unexplored. Third,
\emph{schema evolution}: diff and merge require compatible schemas, yet
schema changes are common in practice, so schema must eventually be
versioned together with data. Fourth, \emph{retention}: named snapshots pin
immutable objects, so sustained branching accumulates history, and balancing
auditability against storage growth and compaction effectiveness requires
explicit policy. Finally, we plan to validate Git4Data on production agentic
workloads to learn how far these lessons.


\balance
\bibliographystyle{ACM-Reference-Format}
\bibliography{paper}

\end{document}